\documentclass[pra,twocolumn,showpacs,preprintnumbers,amsmath,amssymb]{revtex4}
\usepackage{graphicx}
\usepackage{dcolumn}
\usepackage{bm}
\usepackage{trfsigns}
\newcommand \be{\begin{eqnarray}}
\newcommand \ee{\end{eqnarray}}
\newcommand \ba{\begin{align}}
\newcommand {\ket}[1]{|#1\rangle}
\newcommand {\bra}[1]{\langle #1|}
\begin{document}
\title{Universal short-time response and formation of correlations after quantum quenches}
\author{K. Morawetz$^{1,2,3}$}
\affiliation{$^1$M\"unster University of Applied Sciences,
Stegerwaldstrasse 39, 48565 Steinfurt, Germany}
\affiliation{$^2$International Institute of Physics (IIP),
Av. Odilon Gomes de Lima 1722, 59078-400 Natal, Brazil
}
\affiliation{$^{3}$ Max-Planck-Institute for the Physics of Complex Systems, 01187 Dresden, Germany
}
\begin{abstract}
The short-time evolution of two distinct systems, the pump and probe experiments with semiconductor and the sudden quench of cold atoms in an optical lattice, is found to be described by the same universal response function. This analytic formula at short time scales is derived from the quantum kinetic theory approach observing that correlations need time to be formed. The demand of density conservation leads to a reduction of the relaxation time by a factor of four in quench setups.  The influence of finite trapping potential is derived and discussed as well as Singwi-Sj{\o}lander local field corrections. 
\end{abstract}

\maketitle

\section{Introduction}

The experimental progress with cold atoms triggered by Bose-Einstein condensation has led to an enormous activity to understand the time-dependent formation of correlations. Besides pump-and probe experiments in semiconductors like GaAs 
\cite{HTBL02} or InP 
\cite{HKTBLVHKA05} where  
the  formation of collective modes and quasiparticles have been observed with the help of femtosecond spectroscopy,
it is now possible to measure the time-dependent occupation of Hubbard-like set ups and to observe the formation of correlations \cite{TCFMSEB12}.

The ultrafast excitations in semiconductors, clusters, or 
plasmas by ultrashort laser pulses are characterized by long-range Coulomb interactions reflected in the time 
dependence of the dielectric function 
\cite{HTBBAL01,HTBL02,HKTBLVHKA05} in the terahertz 
regime \cite{HCFJJ04}, for an overview over theoretical 
and experimental work see \cite{AK04}. 
Calculating nonequilibrium Green's functions \cite{BVMH98,GBH99} 
allows one to describe the formation of 
collective modes \cite{VH00,HKTBLVHKA05}, screening \cite{BVMH98} and even exciton population inversions \cite{KK04}. 

The thermalisation of cold atoms is characterized by short-range interactions mostly described within Hubbard-like models \cite{GHL12}. Different kinds of quenches are applied and studied \cite{SCC02,MWNRM09} and the importance of local conservation laws has been pointed out \cite{FE13}. Concerning these conservation laws several existing many-body approximations on the market have been analyzed by numerical solution of Kadanoff and Baym equations \cite{FPVA09} and recently \cite{HB14}. Special preparation of cold atoms in optical lattices allows to study the local relaxation \cite{FCMSE08,TCFMSEB12} and to explore dissipation mechanisms \cite{Sya08}. At intermediate time scales a quasistationary state had been found during thermalization process \cite{Eck09}.

Both different physical systems, the long-range Coulomb \cite{MLS05} as well as short-range Hubbard systems can be described by a common theoretical approach leading even to a unique formula to describe the formation of correlations at short-time scale  as we will demonstrate. This could be of interest since normally the  formation is explained by numerically demanding calculations solving Green functions \cite{BVMH98,GBH99} or renormalization group equations \cite{TCFMSEB12}.

What is the basic idea that such different systems show a universal feature? It is just the fact that correlations need time to be formed. In other words higher-order correlations need more time to be build up than low-order ones. Though this statement is strictly valid only for weakly correlated systems, we adopt it here to see that it leads to good results at short-time scales even in the strong-correlated regime. Therefore it is suggested here that the lowest level, the meanfield approximation, is sufficient to describe the basic features of short-time formation of correlations.
We will follow a relaxation-time approximation with explicitly imposing local conservation laws where the meanfields are time-dependent which is an extension of the idea of Mermin \cite{Mer70,D75}. In order to estimate the influence of higher-order correlations on the response, the time-dependent local field corrections have been derived on the level of Singwi-Sj{\o}lander approximation. 

The outline of the paper is as follows. In the next chapter we sketch shortly the kinetic equations in mean-field and conserving relaxation-time approximation and linearize the solution. The influence of a harmonic trap is derived exactly. In the this chapter we solve the kinetic equation for the situation of sudden quench of atomic lattices like Hubbard models leading to an analytical formula describing the time-dependence of the occupation and compare them with the experimental and renormalization group data. This formula brings the first novel result of the paper. The influence of the trapping potential is discussed in the regime of the experiments which allows a perturbative treatment. The fourth chapter is devoted to the response function if initially the system is uncorrelated. Here it is found that the short-time response has the same form for long-range Coulomb interactions and for short-range Hubbard interactions. This universal time-dependence is the second novel result of the paper and the sum rules are discussed and the local-field corrections as expression of higher-order correlations are derived. The time-dependence of the response function for different approximations are illustrated by complimentary movies online. A summary tries to encourage the usage and further comparison with other approaches.

\section{Kinetic theory approach}
The advantage of the equation-of-motions is that the linearization leads to a higher-order response function than used as approximation in the kinetic equation. This means e.g. that a meanfield equation leads to Random Phase Approximation (RPA) response etc which has been pointed out in \cite{KB00,Mc02}. Therefore we will present shortly the ingredients of such kinetic approach and restrict to the meanfield as lowest order since higher-order correlations need time to be formed if one starts uncorrelated \cite{MBMRK01,KM01}.

We consider models with the Hamiltonian
\be
H=\sum\limits_k \epsilon_k a_k^+a_k +\frac 1 2 \sum\limits_{kpq} V_q a_k^+a_p^+a_{p+q}a_{k-q}
\label{H}
\ee 
where the energy dispersion of quasiparticles in Coulomb systems like semiconductors is given by an effective mass $\epsilon_p=p^2/2m$. As model for the short-range lattices we consider the Hubbard model of constant hopping $J$ and constant Coulomb repulsion $V_q=U$. For short-range lattices of length $a$ and with hopping element $J$ the dispersion is given by $\epsilon_p=2 J(1-\cos{a p/\hbar})$ such that $1/m\approx 2 J a^2/\hbar^2$ near the band minima. 

We consider the time evolution of the reduced density matrix $\bra {p+\frac 1 2 q} \delta \rho\ket{p-\frac 1 2 q}=\delta f(p,q,t)$ which is given by linearization $\delta [H,\rho]=[\delta H,\rho_0]+[H_0,\delta \rho]$ of the kinetic equation
\be
\dot {\rho}+i[H,\rho] ={\rho^{\rm l.e.}-\rho \over \tau}
\label{1}
\ee
with respect to an external perturbation $\delta V^{\rm ext}$. The effective Hamiltonian consists of the quasiparticle energy, the external and induced meanfield $\bra {p+\frac 1 2 q}\delta H\ket{p-\frac 1 2 q}=\delta V^{\rm ext}+V_q \delta n_q$ given by the interaction potential $V_q$ and the density variation $\delta n_q$. As possible confining potential we assume a harmonic trap $V^{\rm trap}=\frac 1 2 K x^2$ which leads to $\bra {p+\frac 1 2 q} \delta [V^{\rm trap},\rho] \ket{p-\frac 1 2 q}=-K\partial_p\partial_q \delta f(p,q,t)$.

The kinetic equation (\ref{1}) relaxes towards a local equilibrium (Fermi/Bose) distribution but with an allowed variation of the chemical potential 
$
\bra{p+\frac {q}{ 2}} \rho^{\rm l.e.}-\rho\ket{p-\frac {q} {2}}=\bra{} \rho^{\rm l.e.}-\rho^0\ket{}-\delta f(p,q,t)=-{\Delta f \over \Delta \epsilon}\delta \mu(q,t)-\delta f(p,q,t).
$
Here we use the short-hand notation $\Delta f=f_0({p+\frac q 2})-f_0({p-\frac q 2})$ and $\Delta \epsilon=\epsilon_{p+\frac q 2}-\epsilon_{p-\frac q 2}$. 
The variation of the chemical potential is determined by the density conservation $n=\sum_p f=\sum_p f^{\rm l.e.}$ leading to a relation between density variation $\delta n(q,t)=\tilde \Pi(t,\omega=0) \delta \mu(q,t)$ and the polarization in RPA  
\be
\Pi(t,t')\!&=&\!i\! \sum\limits_p [f_{p\!+\!
\frac q 2}(t')\!-\!f_{p\!-\!\frac q 2}(t')]
{\rm e}^{\left (i \varepsilon_{\!\!p\!+\!\frac q 2}-i 
\varepsilon_{\!\!p\!-\!\frac q 2}\!+\!\frac{1}{\tau}\right )
(t'\!-\!t)},
\nonumber\\
\tilde \Pi(t,\omega)&=&\int d(t-t') {\rm e}^{i\omega (t-t')} \Pi(t,t').
\label{pi}
\ee

The linearized kinetic equation (\ref{1}) reads therefore for $\delta f_t=\delta f(p,q,t)$
\ba
\dot {\delta f}_t+{\delta f_t\over \tau} +i\Delta \epsilon \delta f_t=&i\Delta f \delta V^{\rm ext}_t+i\Delta f V_q\delta n_t\nonumber\\
+&{\Delta f \over \Delta \epsilon} {\delta n_t\over \tau \Pi(q,0,t)}+iK\partial_p\partial_q \delta f_t.
\label{lin}
\end{align}
The last term on the second line describes the confining harmonic trap and the first term comes from Mermin's correction due to density conserving relaxation time approximation.

Neglecting the 
time derivative of the homogeneous part $f_0(p,t)$ compared to $\delta f(p,q,t)$ we can solve this kinetic equation (\ref{lin}) considering the momentum derivatives of the last term as perturbation to obtain
\ba
&\delta f(p,q,t)-\delta f(p,q,0)=i\!\int\limits_{t_0}^t \! dt' 
\exp{\left [\left (i \Delta \varepsilon\!+\!\frac{1}{\tau}\right )(t'-t)\right ]}
\nonumber\\
&
\times
\Biggl \{
\Delta f(t')
\left [V_q \delta n(q,t')+
V_q ^{\rm ext}(t')\right ]
\nonumber\\
&
+{1 \over i \tau \tilde \Pi(t',0)}
{\Delta f(t') \over 
\Delta \varepsilon}
\delta n(q, t')
+K\partial_p\partial_q \delta f(p,q,t')
\Biggr \}. 
\label{f}
\end{align}
The further evaluation is very much dependent on the physical setup and leads to different solutions.

\section{Atoms in lattice after sudden quench}
For cold atoms occupying each second place on a lattice $(1+(-1)^k) n/2$ we have a Fourier transform to the momentum distribution
\be
f_0(p)&=&a\sum\limits_{k=-N}^N {\rm e}^{{i\over \hbar} ka p}f_k= n a {\sin{(2 N+1) {q p\over \hbar}}\over \sin ({a p\over \hbar})}
\nonumber\\
&\to& \pi \hbar n\delta (p)=\frac n 2 \delta_p,
\label{f0}
\ee
for large total number of atoms $N$ and the lattice spacing $a$. We can now Laplace transform the time $t\to s$ in the kinetic equation (\ref{lin}) or (\ref{f}) to get
\ba
&\delta \!f_s\!=\!{\delta f_0\over s\!+\!i\Delta \epsilon\!+\!\!{1\over \tau}}\!+\!\frac {in}{2} \!\!\left (\!{\delta_{p\!+\!q/2}\over s\!-\!i b\!+\!\!{1\over \tau}}\!-\!{\delta_{p\!-\!q/2}\over s\!+\!i b\!+\!\!{1 \over \tau}}\!\right )\! (\delta V^{\rm ext}\!\!+\!V_q\delta n_s\!)\nonumber\\
&+\!\frac{1}{2\tau}\left ({\delta_{p+q/2}\over s\!-\!i b\!+\!{1\over \tau}}\!+\!{\delta_{p-q/2}\over s\!+\!i b\!+\!{1 \over \tau}}\right )\delta n_s\!+\!{i K \over s\!+\!i\Delta \epsilon\!+\!{1\over  \tau}}\partial_p\partial_q \delta f_s
\label{fs}
\end{align}
where we introduced $\Delta \epsilon|_{p=\pm q/2}=\pm 4 n J \sin^2{a q\over 2\hbar}=\pm b$.
The initial time disturbance of the distribution $\delta f_0$ is determined according to the different physical preparations.

In case of a sudden quench the interaction is switched on suddenly and no external perturbation will be assumed $\delta V^{\rm ext}=0$. Lets consider the time evolution of an empty place in the lattice if each second place was initially populated. The density $n_t=\frac n 2 +\delta n_t$ starts with $n_0=0$ which means $\delta n_0=-n/2$ as initial condition. If we first look at the quench without interaction ($V=0$) we can solve (\ref{lin})
\be
\delta f_t=\delta f(0) {\rm e}^{-i\Delta \epsilon t-{t\over \tau}}
\ee
with the choice of $\delta f(0)=-n/2$ that the density 
\be
\delta n_t=\sum\limits_p \delta f_t=
-{n\over 2} J_0(\sqrt{4 J b}t){\rm e}^{-{t\over \tau}}
\ee
starts with $\delta n_0=-n/2$ as desired since the Bessel function $J_0(0)=1$. Please note that $\sum_p ={a/2\pi}\int_0^{2\pi/a} dp $. Now that the initial value is  specified we can integrate (\ref{fs}) over momentum including the interaction to get an equation for the density. Using $J_0(\sqrt{4 J b}t)\Laplace 1/\sqrt{s^2\!+\!4 J b}$ lets first inspect the solution without a confining trap ($K=0$)
\ba
\delta n_s=&-\frac n 2 {(s+\frac 1 \tau)^2+b^2\over \sqrt{(s+\frac 1 \tau )^2\!+\!4 J b}(s^2+\frac s \tau+n b V_q+b^2)}\laplace
\nonumber\\
\delta n_t=&-\frac n 2 J_0(\sqrt{4 J b}t){\rm e}^{-{t\over\tau}}
-\frac {n } {4\gamma \tau^2} \int\limits_0^t d x J_0(\sqrt{4 J b}x){\rm e}^{-{t+x\over 2 \tau}}\nonumber\\
&\times \left (2 \gamma \tau \cos{\gamma (t-x)}+(1-2 b n V_q \tau^2) \sin{\gamma (t-x)}\right )
\label{dnt}
\end{align}
where $\gamma^2=n b V+b^2-1/4\tau^2$.
Without interaction $V=0$ and damping $1/\tau\to 0$ we obtain the exact result of \cite{FCMSE08}.

\begin{figure}[h]
\includegraphics[width=7cm]{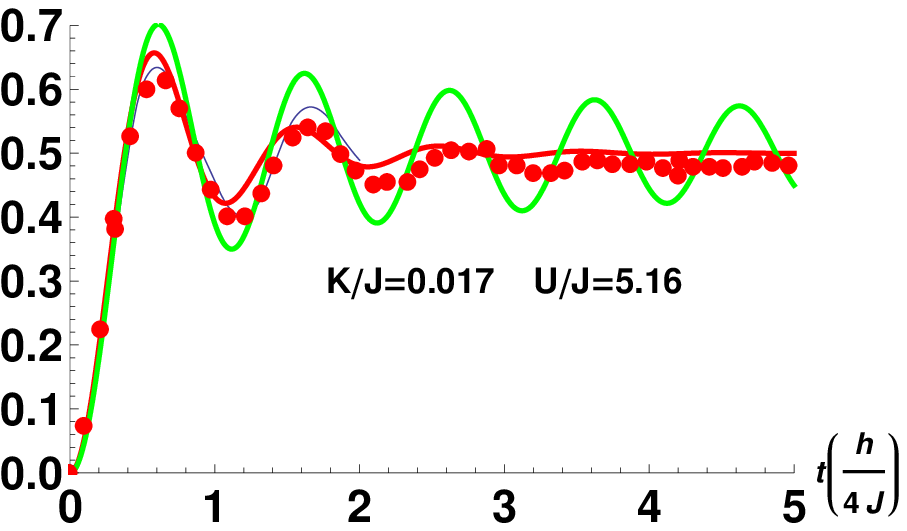}
\includegraphics[width=7cm]{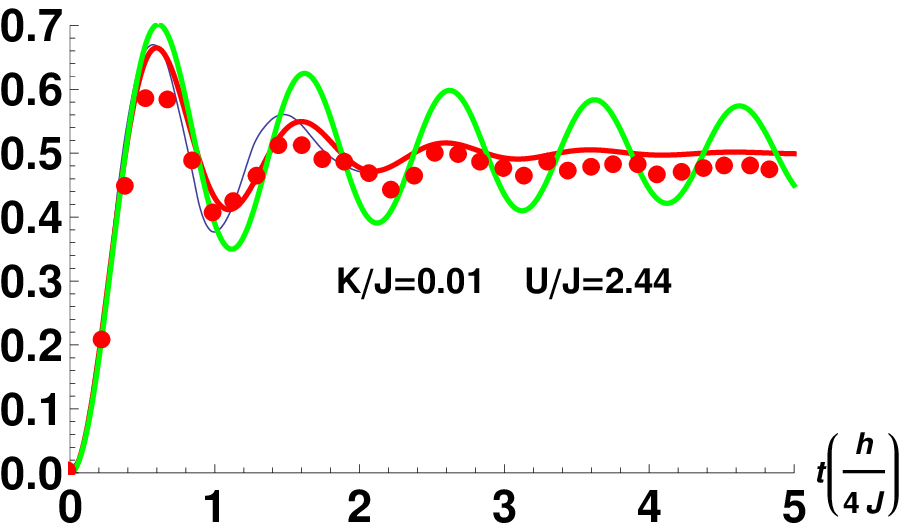}
\includegraphics[width=7cm]{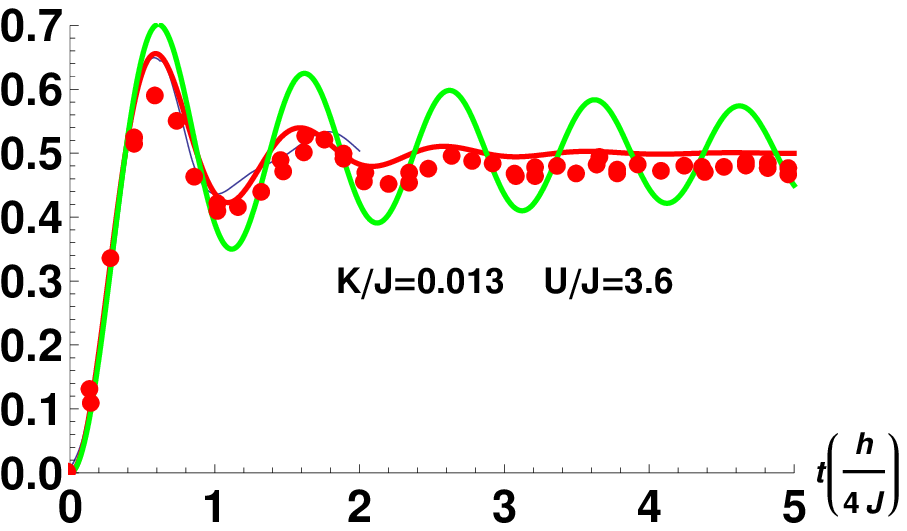}
\includegraphics[width=7cm]{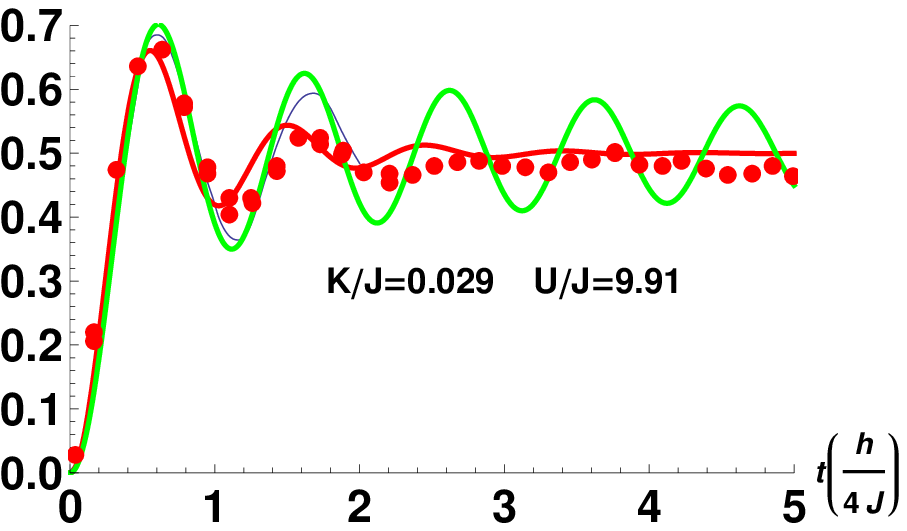}
\caption{Comparison of the experimental data of \cite{TCFMSEB12} (dots) with the RG calculation (thin line) \cite{FCMSE08} and Mermin's correction of conserving relaxation time $\tau=0.6 \hbar/J$ approximation (\ref{dnt})
  without (green) and with interaction (red).
}
\label{bessV}
\end{figure}

In figures \ref{bessV} we compare the experimental data \cite{TCFMSEB12} with (\ref{dnt}) where we plot the interaction free evolution together with the interaction one. The main effect of interaction is the damping which brings the curves nearer to the experimental data. 
Here we use the parameter
for the lattice constant given by half of the short laser wavelength
$a=\lambda/2=765$nm which provides a wave vector of $q=\pi \hbar /a$, 
and an initial density $n=1/2a$ with each second
place filled. The relaxation time characterizes dissipative processes which we assume to arise due to polaron scattering. These lattice deformation processes are dominated by hopping transport at high temperatures and band regime transport at low temperature with the transition given by $\hbar/\tau =2 J \exp{(-S)}$ where $S$ describes the ratio of polaron binding to optical phonon energy. This quantity is generally difficult to calculate \cite{JM73} but in the order of one. We will use it as fit parameter and find a common value $\tau=0.6 \hbar/$ for the results in the figures presented here. 

\begin{figure}
\includegraphics[width=7cm]{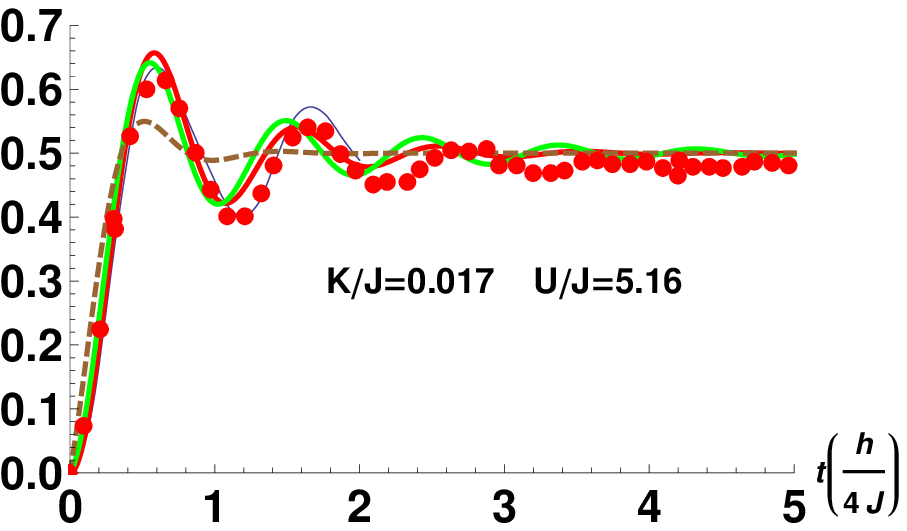}
\includegraphics[width=7cm]{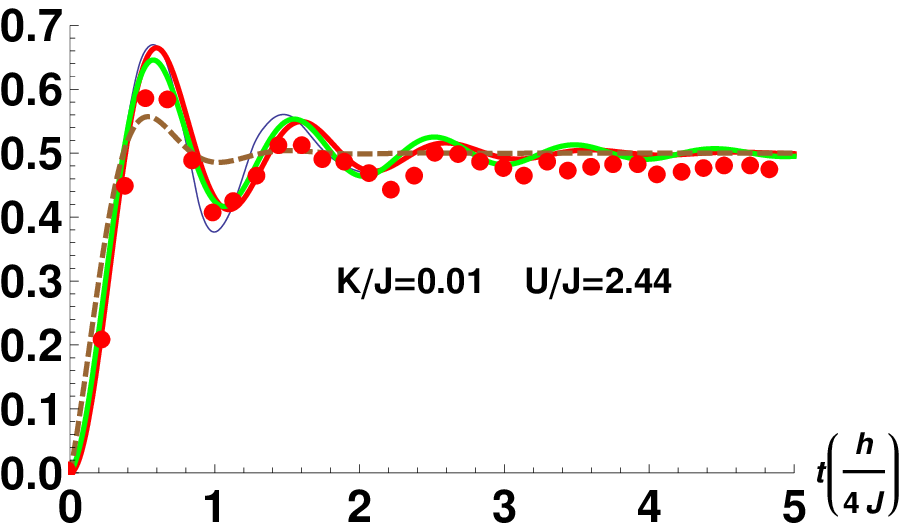}
\includegraphics[width=7cm]{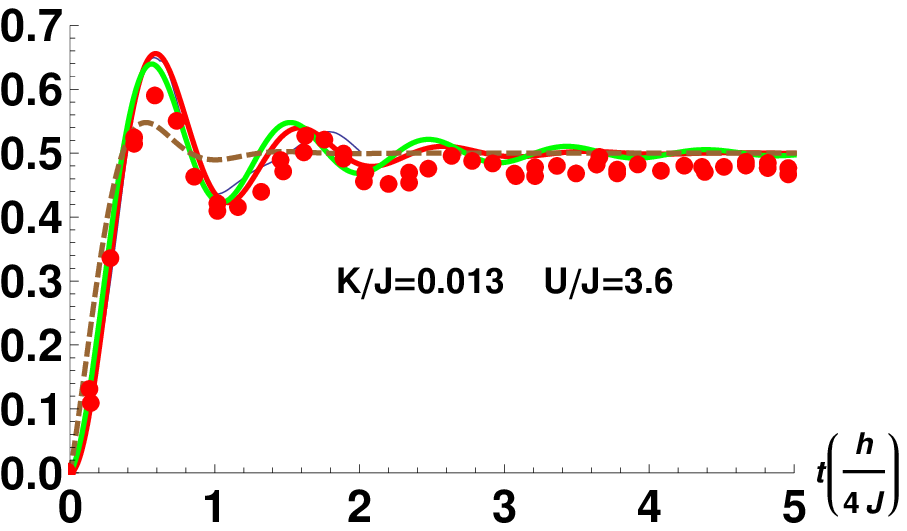}
\includegraphics[width=7cm]{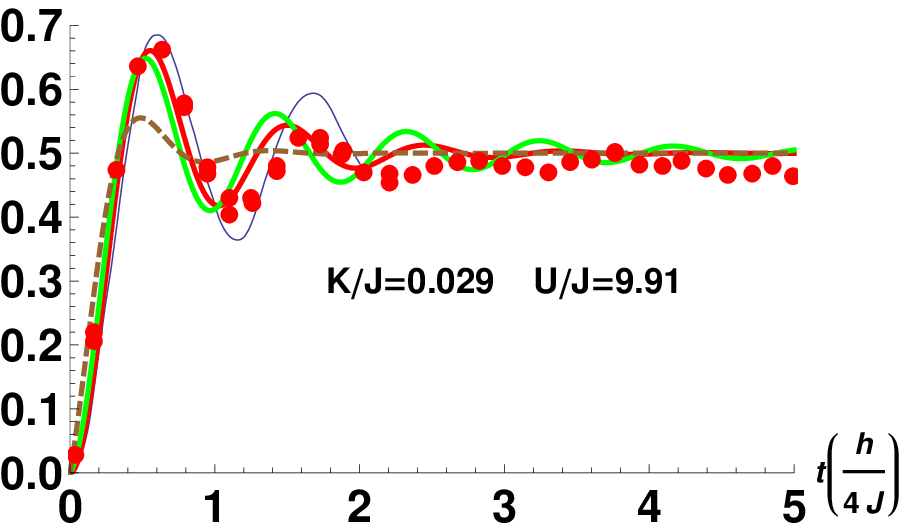}
\caption{Same data as in figure \ref{bessV} and (\ref{dnt}) with (red) and without (dashed brown) Mermin's correction of conserving relaxation time approximation. The curve without Mermin's approximation but 4 times larger relaxation time is plotted as green curve for comparison. 
}
\label{bessV_merm}
\end{figure}

The influence of Mermin's correction is visible in figure \ref{bessV_merm}. One sees that without these corrections of conserving relaxation time approximation (brown curve) we obtain too much damping. In order to compare the relative forms of the curve we increase artificially the relaxation time by a factor 4 (green) which brings both curves nearer together. One sees that without Mermin's correction (and 4 times larger relaxation time) we have a too fast oscillation compared to the data. Mermin's correction as expression for the density conservation diminishes the frequency in better agreement with the data. 

However it is still not sufficient since the oscillations are still too fast. We need the corrections of the finite trap condensed in the parameter $K/J$ which we have not considered yet.

The finite size term $\sim K/J$ is a small number such that one iteration of (\ref{fs}) is sufficient. With one partial integration with respect to $p$ we obtain 3 terms. The first one 
\be
{i K n\over 2}\sum\limits_p\partial_p\left ({1\over s+i\Delta \epsilon+\frac 1 \tau }\right )\partial_q\left ({1\over s+i\Delta \epsilon+\frac 1 \tau }\right )
\ee
is easily seen to vanish by back transforming to time and performing the momentum integration.
The second term is the $q$-derivative of $\delta n$ which is readily integrated due to the $\delta_p$ functions to yield
\be
&&\frac K 2 \partial_q \delta n_s 4 J a \cos{q a\over 2}\sin{q a\over 2}\nonumber\\
&&\times \left (n b V \partial^2_{\frac 1 \tau}-\frac 1 \tau \partial^2_{\frac 1 \tau}(s+\frac 1 \tau)\right ) {1\over (s+\frac 1 \tau)^2+b^2}.
\ee
Since the experimental data are performed with commensurate wavelengths $q=\hbar/a$ this term is negligible small.

\begin{figure}
\includegraphics[width=7cm]{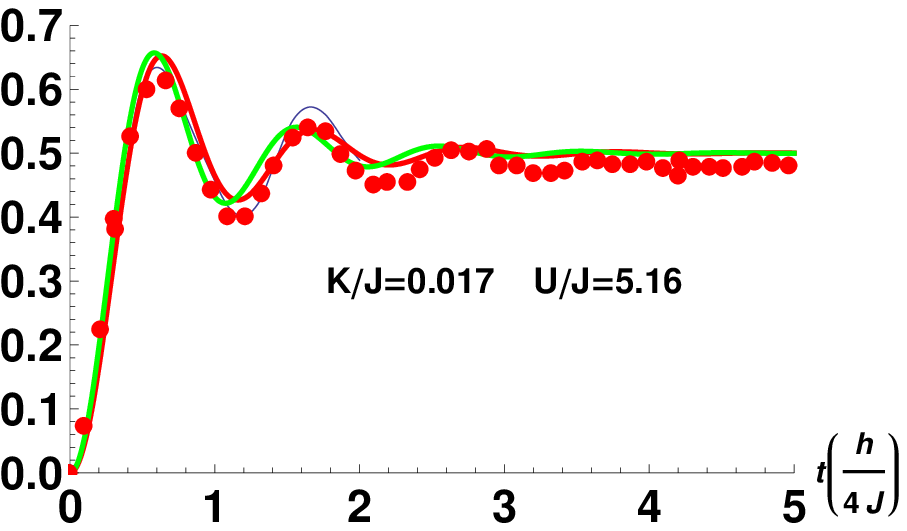}
\includegraphics[width=7cm]{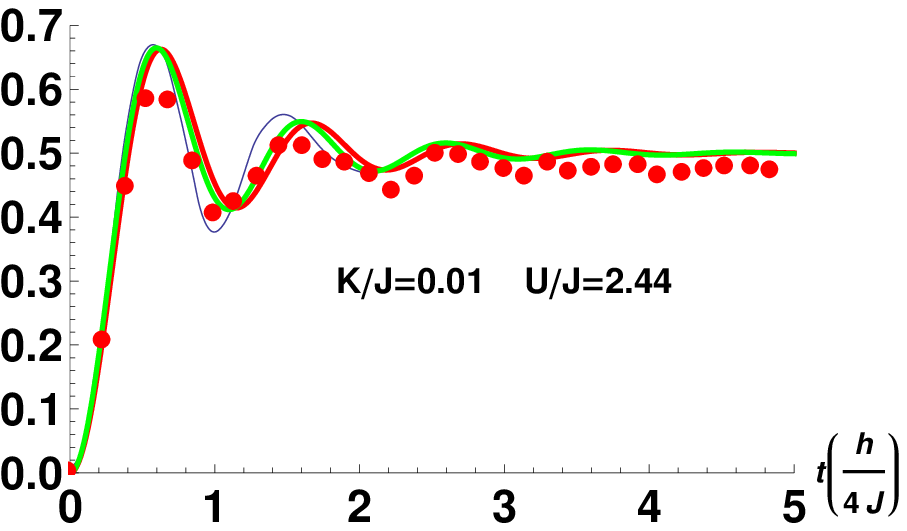}
\includegraphics[width=7cm]{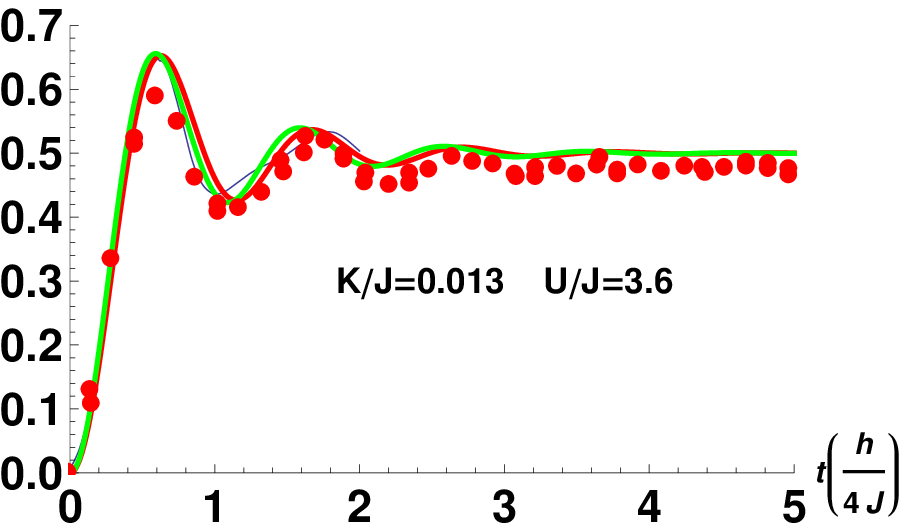}
\includegraphics[width=7cm]{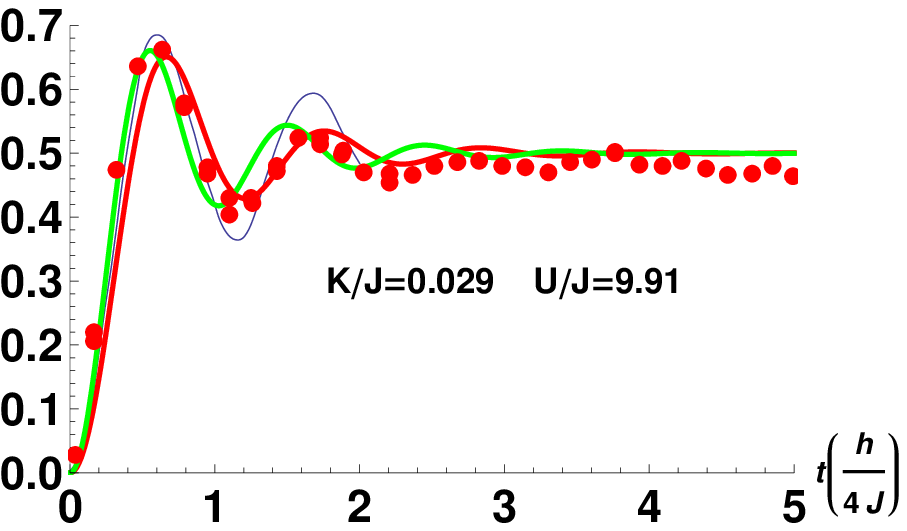}
\caption{Same data as in figure \ref{bessV} and with (red) and without (green) the influence of the trapping potential $K/J$ of (\ref{k}) together with the RG calculation (thin line) of \cite{FCMSE08}.
}
\label{bessV_finite}
\end{figure}

The last remaining term is the one proportional to $\delta n_s$ which results in
\be
&&\frac K 2 \delta n_s 2 J a^2 \sin^2{q a\over 2}
\nonumber\\
&&\times \left (-n b V \partial^2_{\frac 1 \tau}+\frac 1 \tau \partial^2_{\frac 1 \tau}(s+\frac 1 \tau)\right ) {1\over (s+\frac 1 \tau)^2+b^2}
\ee
where we again have neglected terms $\sim \cos {q a\over 2}$. We see that the trap potential introduces an additional $s$ and therefore time dependence. However these correction range from zero at zero time $(s\to\infty)$ to the maximal value at large times $(s\to 0)$.  Using the latter limit we obtain a constant shift to the term $nbV$ in (\ref{dnt}) of
\ba
&nbV \to nbV +k\nonumber\\
&k\!=\!{K a^2b\over 2 (1\!+\!\tau^2b^2)^2}[n b V \tau^2 (\tau^2 b^2\!-\!3)\!-\!1\!+\!3 b^2\tau^2].
\label{k}
\end{align}
In figures \ref{bessV_finite} we see that this trap potential corrections decrease the frequency further and (\ref{dnt}) agrees better with the data though the corrections are small. As comparison we plotted the RG simulation \cite{FCMSE08} as thin lines. For larger interaction $U/J$ we see that the analytic result (\ref{dn}) describes the data slightly better and we can give the time evolution up to more oscillations then possible by numerical RG.

\section{Short-time response function}

\subsection{RPA modes}
Now we are interested in the short-time response of the system to an external perturbation $V^{\rm ext}$. This is different from sudden quench since here we have initially $\delta f(p,q,0)=0$ and the system is driven out of equilibrium by $V^{\rm ext}$. As the result we will obtain the dielectric response which gives microscopic access to optical properties.

Integrating (\ref{f}) over momentum one obtains the time-dependent density response 
\be
\delta n(q,t)=\int\limits_{t_0}^t dt' \chi(t,t')V_q ^{\rm ext}(t')
\label{dn}
\ee
describing the response of the system with respect to the external field in contrast to the polarization function (\ref{pi}) which is the response to the induced field.

One obtains the equation for $\chi(t,t')$ from (\ref{dn}) by interchanging integrations in $(\ref{f})$
\ba
\chi(t,t')\!=\!\Pi(t,t')\!+\!\!
\int\limits_{t'}^t \! \! d {\bar t}\biggl \{ 
\left [ \Pi(t,{\bar t}) V_q \!+\! 
I(t,{\bar t})\right ] \chi({\bar t},t')\!+\!R(t,\bar t)\biggr \}
\label{chi}
\end{align}
with the polarization (\ref{pi}) and
Mermin's correction
\be
I(t,t')=\sum\limits_p {f_{p\!+\!\frac q 2}(t')\!-\!f_{p\!-\!\frac q 2}(t')\over \varepsilon_{p\!+\!\frac q 2}- \varepsilon_{p\!-\!\frac q 2}}{{\rm e}^{\left (i \varepsilon_{p\!+\!\frac q 2}-i \varepsilon_{p\!-\!\frac q 2}\!+\!\frac{1}{\tau}\right )(t'\!-\!t)}\over  \tau \tilde \Pi^{\rm RPA}(t',0)}.
\nonumber\\
&&
\label{i}
\ee
The confining trap potential leads to a term
 \be
R(tt')=
K\sum\limits_p {\rm e}^{\left (i \varepsilon_{p\!+\!\frac q 2}-i \varepsilon_{p\!-\!\frac q 2}\!+\!\frac{1}{\tau}\right )(t'\!-\!t)}{\partial_p\partial_q}\delta f(p,q,t').
\label{R}
\ee

For cold atoms on the lattice we have obtained already the solution (\ref{fs}) which we can use here with $\delta f(0)=0$ and we have
\ba
\Pi(t,t')&=n 
{\rm e} ^{t'-t\over \tau}\sin{[b (t'-t)]}\nonumber\\
I(t,t')&={1\over \tau} 
{\rm e}^{t'-t\over \tau}\cos{[b (t'-t)]}\nonumber\\
R(t,t')&={JKa\over \hbar}\!\int\limits_{t'}^t d t'\! (t'-t) {\rm e}^{(t'-t)\over \tau}{\sin({qa\over 2 \hbar})\partial_q} \chi({\bar t},t').
\end{align}
This will lead to the same response formula as a gas of particles with the thermal Fermi/Bose distribution for $f_p$. 
For the latter one we work in the limit of long wave lengths $q\to 0$ and the leading terms are $\Pi(t,t')\approx {q^2 n(t') \over m} (t'-t) 
{\rm e} ^{t'-t\over \tau}$, $I(t,t')\approx {1\over \tau} 
{\rm e}^{t'-t\over \tau}$, and $R(t,t')=K\int\limits_{t'}^t d t' (t'-t) {\rm e}^{(t'-t)\over \tau}{q\partial_q\over m} \chi({\bar t},t')$ with the time-dependent density $n(t)$. 

We introduce the collective mode of plasma/sound-velocity oscillations for Coulomb gas and for the Hubbard models respectively
\ba
\omega_p^2=\left \{
\begin{array}{lll} {n e^2\over m\varepsilon_0}&{\rm for}& V_q={e^2\hbar^2\over \varepsilon_0 q^2}, \epsilon_p={p^2\over 2m}
\cr
b n a U&{\rm for} &V_q=U a, \epsilon_p=2 J (1-\cos{p a/\hbar})
\end{array}\right .
\end{align} 
where we had used already $b=4 n J \sin^2{a q\over 2\hbar}$. For Coulomb interactions one has an optical mode while for atoms on the lattice the mode is acoustic.

For the gas of particles it is convenient to transform 
(\ref{chi}) into a differential equation
\be
&&\ddot \chi(tt')+\frac 1 \tau \dot \chi(tt')+\omega_p^2\chi(tt')=-k\omega_p^2 \partial_{\omega_p^2} \chi(tt`)+o(k \omega_p^2)\nonumber\\
&& \chi(t,t)=0,\dot\chi(t,t')|_{t=t'}=-{\omega_p^2/V_q}+o(k \omega_p^2)
\label{chi1}
\ee
where the influence of the trap is condensed
in $k=2K/m$ for the gas system and (\ref{k}) for Hubbard models respectively.

\begin{figure}
\includegraphics[width=8.5cm]{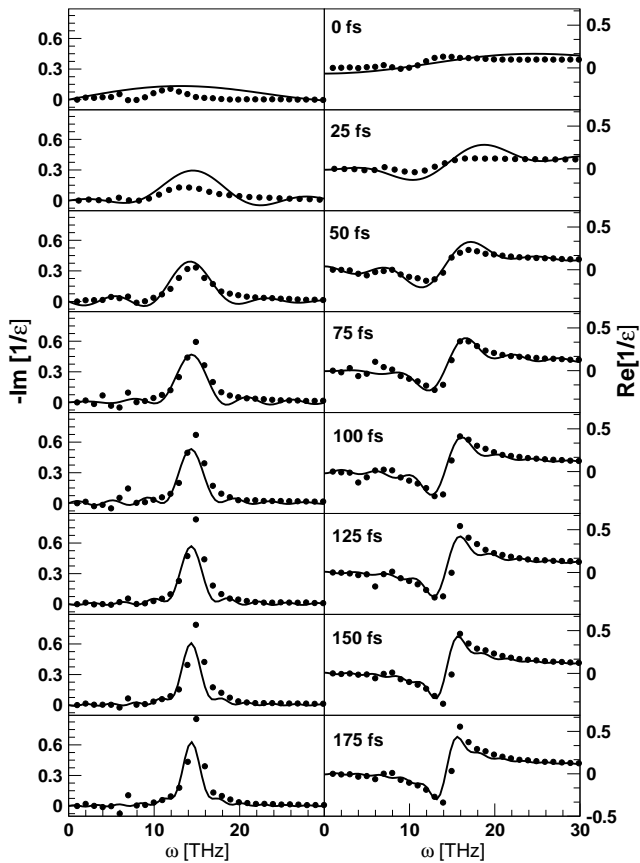}
\caption{The time evolution  
of the inverse dielectric function in 
GaAs. The labels from top to bottom denote the time $t$. The pump pulse was at $t_0=-40$fs and the probe pulse has a full-width at half-maximum of 27$fs$. Circles are data from \protect\cite{HTBBAL01,HTBL02}, 
solid lines show the electronic part (\protect\ref{result}) published in \cite{MLS05}. The plasma frequency is given by $\omega_p=$14.4 THz and the relaxation time is $\tau=85$ fs.  
}
\label{GaAs_1}
\end{figure}

Interestingly, both solutions, the one for the Hubbard lattice (\ref{fs}) and the one for the gas of particle (\ref{chi1}) lead to the same result of the integral equation (\ref{chi}) for the two-time response function
\be
V \chi(t,t')=-{\omega_p^2\over\gamma} 
{\rm e}^{-{t-t'\over 2 \tau} } \,\sin{\gamma (t-t')}
\label{result}
\ee
but with a different collective mode $\gamma=\sqrt{\omega_p ^2+k-{1\over 4 \tau ^2}}$ for the Coulomb gas and $\gamma=\sqrt{\omega_p ^2+b^2+k-{1\over 4 \tau ^2}}$ for cold atoms. 
In this sense we consider (\ref{result}) as universal short-time behavior.

In the further analysis we will follow closely the experimental way of analyzing the two-time response function \cite{HKTBLVHKA05}. The pump pulse is creating charge carriers in the conduction band and the probe pulse is testing the time evolution of this occupation. The time delay after this probe pulse  $T=t-t_0$ is Fourier transformed into frequency. Similarly we start the half empty lattice of cold atoms to relax at $t_0$. The frequency-dependent inverse dielectric 
function associated with the actual time $t$ is then given by
\be
{1 \over \varepsilon(\omega,t)}=1+
\int\limits_{0}^{t-t_0} dT {\rm e}^{i \omega T}  V\chi(t,t-T)
\label{ft}
\ee
which is exactly the one-sided 
Fourier transform introduced in Ref. \cite{SBHHH94}. 
The integral (\ref{ft}) with (\ref{result}) can be expressed in terms of elementary 
functions.
Without the last term due to the confining potential 
it is the solution derived for Coulomb systems in \cite{MLS05}.

The virtue of (\ref{ft}) is that the long-time limit 
yields correctly the Drude formula ($K\to 0$)
\be
\lim\limits_{t\to\infty} {1 \over \varepsilon}=
1-{\omega_p ^2\over \gamma ^2-
\omega(\omega+{i\over \tau})}
\label{tinf}
\ee    
which is not easy to achieve within short-time expansions \cite{SBHHH94} and which had provided 
the wrong long-time limit $1-\omega_p^2/[\omega_p^2-(\omega+i/\tau)^2]$ before. 

\subsection{Sum rules}
Checking on the sum rules we first see that the f-sum rule is completed independent on time. Indeed using $\int_0^\infty \omega \sin{\omega T}=-\delta' (T) \pi$ we obtain from (\ref{ft})
\be
\int\limits_0^\infty \omega {\rm Im} \varepsilon^{-1}=-\pi \omega_p^2.
\ee 

The compressibility sum rule becomes time dependent and reads
\be
&&\int\limits_0^\infty {d\omega \over \omega} {\rm Im} \varepsilon^{-1}=-{\pi\omega_p^2\over 2 \gamma} \int\limits_0^{t-t_0}dT {\rm e}^{-{T\over 2 \tau}}{\sin{\gamma T}}\nonumber\\
&\to& 
-{\pi\over 2}{\omega_p^2\over \gamma^2+k}=
-{\pi\over 2}\left \{\begin{array}{l}
{\omega_p^2\over \omega_p^2+k}
\cr
{\omega_p^2\over \omega_p^2+b^2+k}
\end{array}\right .\quad {\rm for}\, t\to\infty
\ee 
for the gas of particles/lattice atoms respectively.
 
This result can be confirmed form (\ref{tinf}) by using the Kramers-Kronig relation
\be
\frac 2 \pi \int\limits_0^\infty {d\omega \over \omega} {\rm Im} \chi&=&
{\rm Re}\chi(\omega=0)=
-{1\over V_q}\left \{\begin{array}{l}
{\omega_p^2\over \omega_p^2+k}
\cr
{\omega_p^2\over \omega_p^2+b^2+k}
\end{array}\right .
\ee
which long wavelength limit $q\to0$ defines the compressibility for the lattice gas
\be
\kappa=\lim\limits_{q\to 0}
{1\over n^2 V_q(\omega_p^2+b^2+k)}
=
{1\over n (n a U-{K a^2/2})}
\ee
where we have used the long-wavelength limit of (\ref{k}).

For the long-ranged Coulomb gas the compressibility is defined as the response to the screened field (E instead of D)
\be
\kappa=-{1\over n^2}\lim\limits_{q\to 0}{\chi(q,0)\over 1+V_q\chi(q,0)}
=\lim\limits_{q\to 0}{\omega_p^2\over n^2 V_q (s^2 q^2+k)}
\ee
where one needs to expand one step further in powers of $q$ in the polarization function to get the sound velocity $s$.
The compressibility vanishes if we have a trap $k\ne 0$. Without trap we have $\kappa=1/nm s^2$ for the gas. 

In spite of the numerous approximations, Eq. (\ref{ft}) with (\ref{result}) 
fits well the experimental data, as shown in Fig.~\ref{GaAs_1}
for the polar semiconductor GaAs. 
The formula (\ref{result}) results in a slightly too fast build-up of the 
collective mode at the time $t=25$ fs. This is, however, just the time duration of the experimental pulse 
and consequently the time of populating the conduction band which we have approximated by an instant jump \cite{MLS05}.

\begin{figure}
\includegraphics[width=8cm]{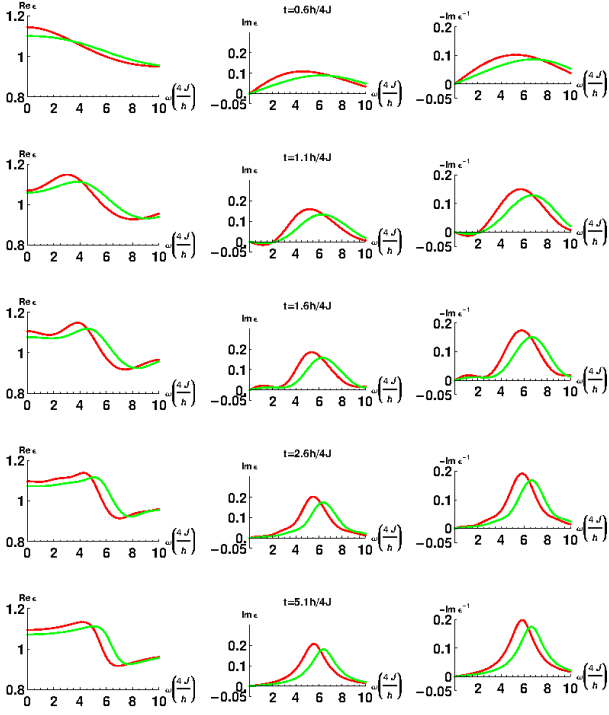}
\caption{The time evolution  
of the dielectric function for the atomic lattice for
$U/J=9.91$ of the figures \ref{bessV_merm}  with (red) and without(green) trap correction according to (\ref{k}) (movie online).
\label{resp_merm}}
\end{figure}

For the atoms on a lattice we obtain the time evolution of the dielectric function as plotted by snapshots in figure \ref{resp_merm}. We see that the influence of finite trap correction diminishes the frequency of the collective mode as we had seen already with the sudden quench. Further the damping is lowered which is visible by a sharpening of the mode. Note the instability around times of $1$h/4J at frequencies of $4$J/h where ${\rm Im}\varepsilon$ becomes negative.

\subsection{Local-field correction}

For strong interactions like the cold atoms feel in the lattice one might expect that the meanfield response is not sufficient. Though it works well at short times since the interactions need time to be formed, for larger times we expect to see the influence of the so called local-field corrections. They describe the interaction cloud around the atom which changes the potential locally. We will derive it here for the lattice atoms in the sense of Singwi-Sj{\o}lander \cite{STLS68}. For this purpose we calculate the second time derivative of the reduced density matrix since this describes the dynamics.  With the help of $i\dot \rho_k=[\rho_k,H]$ the Hamiltonian (\ref{H})
leads to
\ba
\ddot\rho_k=&-\sum\limits_{k_1}(\epsilon_{k_1+k}-\epsilon_{k_1})^2 a^+_{k_1}a_{k_1+k}\nonumber\\
&-\frac 1 2 \sum\limits_{k_1 q} V_q (\epsilon_{k_1+k}-\epsilon_{k_1}-\epsilon_{k_1+k-q}+\epsilon_{k_1-q})
\nonumber\\
&\qquad\qquad\times(\rho_q a_{k_1}^+a_{k_1+k-q}+a_{k_1}^ +a_{k_1+k-q}\rho_q).
\end{align}
The trick is now to write the meanfield term $k=q$ in the sum of the second term in front of the sum. For lattice atoms with the dispersion $\epsilon_p=2 J (1-\cos{p a})$ one obtains
\ba
&\ddot {\rho_k}=-4 b^2\sum\limits_{k_1} \left ({\sin{k_1 a\over 2}\over\sin{k a\over 2}}\right )^2 \cos^2{k_1 a\over 2} f_{k_1,k}
\nonumber\\
&-n b V_k
\sum\limits_{k_1}\cos{(k_1 a)}\left (\rho_k f_{k_1,0}+f_{k_1,0} \rho_q\right)
\nonumber\\
&\times\left [1\!+\!{
\sum\limits_{q}{V_q\sin{q a\over 2}\over V_k\sin{k a\over 2}}
\sum\limits_{k_1}\cos{(k_1 a)} 
\left (\!\rho_q f_{k_1,k-q}\!+\!f_{k_1,k-q}\rho_q\!\right)
\over
\sum\limits_{k_1}\cos{(k_1 a)} \left (\rho_k f_{k_1,0}+f_{k_1,0}\rho_k\right )
}\right ]
\label{local}
\end{align}
and one sees that the effect of correlations beyond the meanfield can be recast into a local field $V_k\to V_k[1+G_k(t)]$. Introducing the $cos$ weighted density $\rho^c_q=\sum_k\cos{(ka)}f_{k,q}$ this local field is read-off (\ref{local})
\be
G_k=\sum\limits_{q}{V_q\sin{q a\over 2}\over V_k\sin{k a\over 2}} 
{\rho_q \rho^c_{k-q}\!+\!\rho_{k-q}^c\rho_q\over\rho_k \rho^c_{0}\!+\!\rho^c_{0}\rho_k}.
\ee

We linearize now the Wigner function $f_{p,q}=a^+_{p-\frac q 2}a_{p+\frac q 2}=\frac n 2 \delta_p+\delta f_{p,q}$ according to the special equilibrium distribution (\ref{f0}) and the density operator as integral over $p$ correspondingly $\rho_q=\rho_0+\delta \rho_q$. Then $\rho_0=\rho_0^c=n/2$ and we assume $\delta\rho_q=\delta\rho_{-q}$ to arrive for Hubbard models $V_k=U a$ at a wave vector-independent local field
\be
\delta G(t)=\sum\limits_{q}\cos{\left ({q a\over 2}\right)} 
{\delta \rho_q^c(t)\over\rho_0^c}.
\ee

\begin{figure}
\includegraphics[width=8cm]{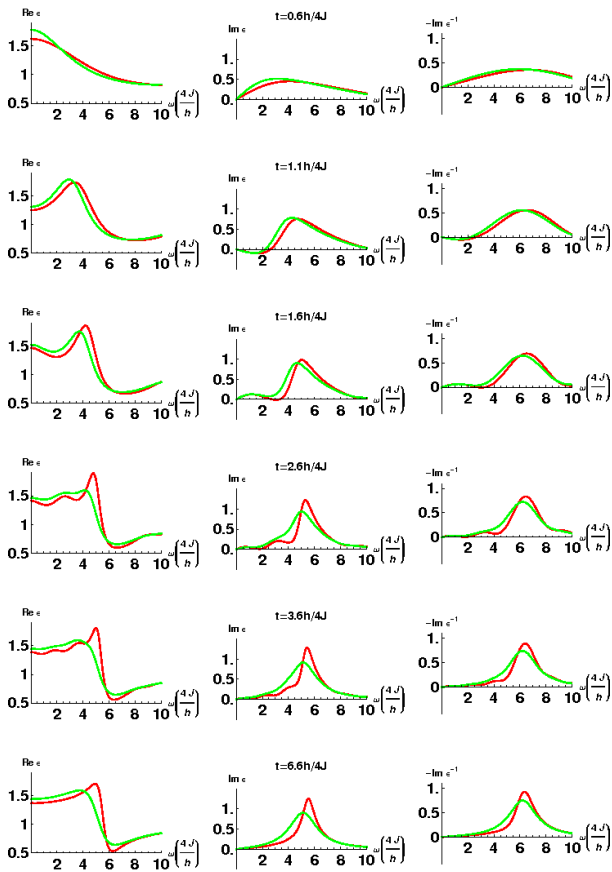}
\caption{The time evolution  
of the dielectric function for the atomic lattice of figures \ref{resp_merm} with
$U/J=9.91$ with (red) and without (green) local field correction of (\ref{gt}) (movie online).
\label{resp_local}}
\end{figure}

This local field correction means that we have to replace in the kinetic equation (\ref{lin}) the term $V_q\delta n\to V_q\delta n+n V_q \delta G$. It is not hard to see from the solution (\ref{f}) that $\delta n^c=\delta n \cos{q a\over 2}$ holds. We obtain for the Laplace transform of the response (\ref{dn}) and (\ref{chi}) which was $\chi_s(k)=-n b(k) /(s^2+\frac s \tau +n b(k) V_k+ b(k)^2)$ the modification
\be
\delta n_s(k)&=&\chi_s(k) \left [
V_s^{\rm ext}(k)+a U \sum\limits_{q}\cos^2{{q a\over 2}} 
\delta n_s(q)
\right ]
\nonumber\\
&=&{\chi_s(k)\over 1-V \delta \bar G_s} V_s^{\rm ext}(k)=\chi_s^{\rm eff}(k) V_s^{\rm ext}(k)
\label{solG}
\ee
where $b(q)=4 J \sin^2{(q a/2)}$. For solving (\ref{solG}) we have integrated the first line with $\cos^2$ leading to
\be
\delta \bar G_s&=&-\sum\limits_{q}{n b(q)\cos^2{\left ({q a\over 2}\right )}\over s^2+\frac s \tau +n b(q) V+ b(q)^2}
\nonumber\\
&=&
-{1\over \pi V} \int\limits_0^1 dy{\sqrt{y-y^2}\over z-y+{4 J\over n V}\, y (y-2)}
\ee
where $z=[(s+{1\over 2\tau})^2+\gamma^2]/4 n V J$ and $\gamma$ of (\ref{result}). Since the response function without local-field correction is just $\chi=-n b/z$ it is advantageous to expand with respect to $1/z$ to obtain
\be
V \chi^{\rm eff}_s=-\sin^2{q a\over 2} \left (
{1\over z}-{1\over 8 z^2}-{3+{22 J\over n V }\over 64z^3}+o(z^{-4})\right ).
\ee
This is easily back-transformed into time
\ba
V\chi_t^{\rm eff}=-\omega_p^2 {\rm e}^{-{t\over 2 \tau}}\left [
1\!+\!{\omega_p^2\over 16 \gamma}{\partial \over \partial_\gamma}
\left (\!1\!-\!{3\!+\!{22 J \over n V}\over 256 \gamma}\omega_p^2{\partial \over \partial_\gamma}\!\right )
\right ]
{sin{\gamma t}\over \gamma}
\label{gt}
\end{align}
which correction is discussed in figure \ref{resp_local}. We see that the local field correction shifts the collective mode towards higher energies and sharpens the mode. Further it leads to more structure at smaller frequencies which heal out at larger times.

\section{Summary}
The aim of the present paper was to separate the gross features of the 
formation of collective modes at transient times which are due to 
mean-field fluctuations. Assuming higher-order correlations described by a conserving relaxation time approximations, this has resulted in an analytic formula 
for the time dependence of the dielectric function and for sudden quench 
dynamics. Subtracting this 
gross feature from the data allows one to extract the effects which are from 
higher-order correlations and which have to be simulated by quantum 
kinetic theory \cite{BVMH98,GBH99,VH00,KK04} and response functions 
with approximations beyond the mean field \cite{KB00}. These treatments are
numerically demanding such that analytic 
expressions for the time dependence of some variables \cite{MSL97a}
are useful for controlling the numerics. We have discussed here also the Singwi-Sj{\o}lander local field correction at short time scales in order to estimate the influence of higher-order correlations at short time behavior. The final answer, however, up to which time the presented short-time expansion works can only be given by solving the time evolution of the complete correlation functions \cite{FPVA09}. Here a selective comparison with experiments and RG simulations in the strong-coupling regime has been chosen as a first test. Further comparisons are needed and encouraged.

As one result of the paper we derive the influence of a finite trapping potential on the kinetic equation and the time evolution of the population. We find that the main effect of correlations consists in lowering the collective frequency and a damping. 

To conclude, we have derived an analytic formula for
the formation of correlations after a sudden quench and for the density response to an external perturbing field. By considering two distinct physical systems, the pump and probe dynamics in semiconductors and the dynamics of atoms in an optical lattice we find the same short-time
feature of the formation 
of quasiparticles which we suggest to be universal. 
The simplicity of the presented
result is extremely practical and offers a wide range of
applications. It could spare a lot of 
computational power to simulate ultrashort-time behavior of new
nano-devices and it can help to understand and describe 
the formation of collective modes during processes which are
not experimentally accessible in the early phase of reactions like in nuclear collisions.

\bibliography{kmsr,kmsr1,kmsr2,kmsr3,kmsr4,kmsr5,kmsr6,kmsr7,delay2,spin,refer,delay3,gdr,chaos,sem3,sem1,sem2,short,quench,spinhall,bose}

\end{document}